\documentclass[doublecol]{epl2} 
 
\title{Efficiency at maximum power of minimally nonlinear irreversible heat engines}

\author{Y. Izumida \and K. Okuda}
\shortauthor{Y. Izumida \etal}

\institute{                    
Division of Physics, Hokkaido University - Sapporo, 060-0810, Japan
}


\abstract{
We propose the minimally nonlinear irreversible heat engine as a new
general theoretical model to study the efficiency at the maximum
power $\eta^*$ of heat engines operating between the 
hot heat reservoir at the temperature $T_h$ 
and the cold one at $T_c$ ($T_c \le T_h $). 
Our model is based on the  
extended Onsager relations with a new nonlinear term meaning the power
dissipation. In this model, 
we show that $\eta^*$ is bounded from the upper side by a function of 
the Carnot efficiency $\eta_C\equiv 1-T_c/T_h$ as
$\eta^*\le \eta_C/(2-\eta_C)$. 
We demonstrate the validity of our theory by showing 
that the low-dissipation Carnot engine can easily be 
described by our theory.}

\begin{document}

\maketitle

\section{Introduction}        
Facing with the recent worldwide problems such as the global warming 
and the depletion of energy resources,    
we have been urged to a        
low-carbon sustainable society. 
Demands for more efficient and greener heat engines have 
rapidly been rising
since heat engines convert heat energy into useful work by 
utilizing only temperature difference, which is   
abundant in the earth's environment:
geothermal power generation, solar thermal power generation, etc. 
might be promising candidates, for example.  
To evaluate and control performance of heat engines, 
it must be important to know the upper bound of    
the energy conversion efficiency of them.  
The efficiency $\eta$ for a heat engine is defined as $\eta \equiv
W/Q_h$, where ${Q}_h$ and $W$ denote the heat transferred from the
hot heat reservoir at the temperature $T_h$ and the work
output, respectively.
Defining $Q_c$ as the heat transferred from the cold heat reservoir at the 
temperature $T_c$ ($\le T_h$), 
we can express $W$ as $W\equiv Q_h+Q_c$.
Thermodynamics tells us that $\eta$ is bounded from the upper side as  
\begin{eqnarray} 
\eta\le 1-\frac{T_c}{T_h} \equiv \eta_C, 
\end{eqnarray} 
where $\eta_C$ denotes the Carnot efficiency and 
the equality holds only when the heat engine is infinitely 
slowly (quasistatically)    
operated to satisfy reversibility.  
Because the heat engine realizing the Carnot efficiency takes infinte time to output a finite amount of work, its power (work output per unit time) is absolutely 0 and thus it is of no practical use.
Motivated by this fact, 
Curzon and Ahlborn~\cite{CA}                                                   
proposed a phenomenological finite-time Carnot cycle model 
and derived that the efficiency
at the maximum power $\eta^*$ of their model is given by an
appealing expression as        
\begin{eqnarray}                                                     
\eta^*=1-\sqrt{\frac{T_c}{T_h}}\equiv \eta_{CA},\label{eq.ca}          
\end{eqnarray}                                                                  
which reminds us of the Carnot efficiency.
Historically and strictly speaking, the formula $1-\sqrt{T_c/T_h}$               
itself was derived by others~\cite{Y,N} 
more previously than~\cite{CA}.                              
But it is usually called the Curzon-Ahlborn (CA) efficiency.  
We also call it the CA efficiency here in accord with the custom.
The paper by Curzon and Ahlborn triggered subsequent 
studies on the efficiency at the 
maximum power of various heat engine models~\cite{R1,R2,LL,G,K,B,AB1,AB2,VB,GS,CH1,CH2,GZC,RB,BK,BJM,AJM,J,REC,ZS,SS,TU,ELB1,ELB2,IO1,IO2,IO3,IO4,GMS,MGS,EKLB1,EKLB2,BSC,US}.
Among recent studies on the CA efficiency, 
it is an important progress that Van den Broeck~\cite{VB}                                         
proved that the CA efficiency $\Delta T/(2T)=\eta_{CA}+O({\Delta T}^2)$              
is the upper bound of the efficiency at the maximum power for the heat engines working                          
in the linear response regime. 
Here, we have defined the temperature difference $\Delta T\equiv T_h-T_c$,
which is assumed to be small, 
and the averaged temperature $T\equiv (T_h+T_c)/2$, respectively.              
Those heat engines working in the linear response regime, which we 
call the linear irreversible heat engines, are described by  
the following Onsager relations~\cite{O,GM}: 
\begin{eqnarray} 
J_1=L_{11}X_1+L_{12}X_2,\label{eq.onsagerJ1}\\  
J_2=L_{21}X_1+L_{22}X_2,\label{eq.onsagerJ2}     
\end{eqnarray}                                                                
where $X_1 \equiv F/T_c \simeq F/T$ with an external force $F$, 
$J_1 \equiv \dot{x}$ with the conjugate variable $x$ of $F$,
$X_2\equiv 1/T_c-1/T_h \simeq \Delta T/T^2$, $J_2\equiv \dot{Q}_h$ and  
$L_{ij}$'s are the Onsager coefficients with the 
reciprocity $L_{12}=L_{21}$.
The dot denotes the quantity per unit time.
Regarding $X_1$ as the control parameter 
for the maximization of the power
$P=-F\dot{x}=-J_1X_1T$,
we can see that the maximum power $P^*$ and the efficiency at the maximum
power $\eta^*$ of the linear irreversible heat engines 
described by eqs.~(\ref{eq.onsagerJ1}) and (\ref{eq.onsagerJ2})
are given by
\begin{eqnarray}
&&P^*=\frac{q^2L_{22}{\Delta T}^2}{4T^3},\\ 
&&\eta^*=\frac{\Delta T}{2T}\frac{q^2}{2-q^2},\label{eq.broecketa}
\end{eqnarray} 
respectively~\cite{VB}, 
where $q$ is called the coupling strength parameter and is defined as 
\begin{eqnarray} 
q\equiv \frac{L_{12}}{\sqrt{L_{11}L_{22}}}.\label{eq.csp}
\end{eqnarray} 
Since the positivity of the entropy production 
rate $\dot{\sigma}\equiv J_1X_1+J_2X_2$, which is a 
quadratic form of the thermodynamic forces, 
restricts the Onsager coefficients $L_{ij}$'s to 
\begin{eqnarray}
L_{11}\ge 0, \ L_{22}\ge 0, \ L_{11}L_{22}-L_{12}L_{21} \ge 0,\label{eq.restriction}
\end{eqnarray}                            
$q$ should take 
$-1 \le q \le 1$. 
Thus $\eta^*$ in eq.~(\ref{eq.broecketa}) takes the upper bound
$\Delta T/(2T)$ which is equal to the CA 
efficiency up to the linear order of $\Delta T$, 
when the tight-coupling condition  
\begin{eqnarray}
|q|=\biggl|\frac{L_{12}}{\sqrt{L_{11}L_{22}}}\biggr|=1\label{eq.tight} 
\end{eqnarray}                        
holds. This condition is equivalent to saying that the two thermodynamic fluxes 
become proportional as $J_2 \propto J_1$.                  
Due to the generality of the theory, the study by Van den Broeck 
renewed the interests in the CA efficiency and inspired 
the studies on the efficiency at the maximum power of 
various heat engine models:
it has been indeed shown that 
$\eta^*$ of many heat engine models,                            
ranging from the steady-state brownian motors~\cite{GS,RB,GZC} to
the finite-time Carnot cycles~\cite{IO3,IO4}, is                       
given by eq.~(\ref{eq.broecketa}) in the linear response 
regime.
Even when the system                                    
begins to enter the nonlinear response regime,                              
it is often observed that 
$\eta^*$ agrees with the CA efficiency 
up to the quadratic order of $\Delta T$ as 
$\eta^*=\Delta T/(2T)+{\Delta T}^2/(8T^2)=\eta_{CA}+O({\Delta T}^3)$~\cite{SS,TU,ELB1,ELB2,EKLB1,EKLB2}.
This fact was firstly observed in~\cite{SS} and proposed as a 
conjecture in~\cite{TU}. Later, it was proved to be a precise result 
for the system which satisfies the left-right symmetry
condition in addition to 
the tight-coupling condition in~\cite{ELB2}.
However, it is also shown, for example 
in~\cite{BJM,AJM,SS,TU,ELB1,ELB2,IO1,IO2,EKLB1,EKLB2}, that 
$\eta^*$ can exceed the CA efficiency 
in the nonlinear response regime,
when those conditions do not hold. 
Therefore the CA efficiency is no longer the upper 
bound of $\eta^*$ for nonlinear irreversible heat engines and 
we need to construct a general theory to determine the upper bound 
of $\eta^*$ for them.

In this paper,    
we propose the minimally nonlinear irreversible 
heat engine described by the extended 
Onsager relations, where a new 
nonlinear term $-\gamma_h {J_1}^2$ meaning the power dissipation 
is added to eq.~(\ref{eq.onsagerJ2}) (see eq.~(\ref{eq.J2})). 
The addition of this new nonlinear 
term can be seen as a natural extension of the 
linear irreversible heat engine and 
we formulate $\eta^*$ of such nonlinear irreversible heat engines. 
Our new formula eq.~(\ref{eq.effi-at-pmax}) contains 
the coupling strength parameter $q$ and a parameter $\gamma_c/\gamma_h$,
where $\gamma_c$ and $\gamma_h$ denote the degree of dissipation to the 
cold and hot heat reservoirs, respectively.
We show that our $\eta^*$ is bounded from the upper
side by a function of the Carnot efficiency $\eta_C$ as 
$\eta^*\le \eta_+$, where
$\eta_+\equiv \eta_C/(2-\eta_C)$. 
Remarkably this $\eta_+$ was also mentioned in the previous 
studies on various finite-time heat engine  
models~\cite{SS,EKLB1,EKLB2,GMS,MGS,CY,VRMC}. 
The generality of our theory allows us to unify these
previous results and explain the universality of $\eta_+$.
For a demonstration of the validity of our theory, we show that a finite-time
Carnot cycle model, called the low-dissipation Carnot engine~\cite{EKLB2}, 
can be described by the extended Onsager relations.      

\section{Extended Onsager relations}
Let us consider that a general heat engine is working  
between the hot and the cold heat reservoirs
with the temperature difference $\Delta T=T_h-T_c$.
Heat engines are generally classified into  
two types: steady-state heat engines and cyclic heat engines. 
Steady-state heat engines literally work in a steady
state since an external force applied on the heat engines 
is time-independent and the 
hot and cold heat reservoirs contact with the heat engines simultaneously. 
Cyclic heat engines, in contrast, work cyclically 
in a time-dependent way such that 
the hot and the cold heat reservoirs contact with the heat engines 
alternately, not simultaneously.  
Our theory below can treat both types of heat engines in a unified manner.
We can generally write
the total entropy production rate $\dot{\sigma}$ of the heat engine as  
the entropy increase rate of the heat reservoirs as 
\begin{eqnarray}
\dot{\sigma}=-\frac{\dot{Q}_h}{T_h}-\frac{\dot{Q}_c}{T_c}
=-\frac{\dot{W}}{T_c}+\dot{Q}_h \left(\frac{1}{T_c}-\frac{1}{T_h}\right),\label{eq.entropy-engine}
\end{eqnarray}
where we do not need to consider the entropy increase inside the heat engine
since the heat engine itself 
is always in a steady state or comes back to 
the original state after one cycle. 
Note that the 
dot denotes the quantity per unit time for steady-state heat engines 
and the quantity divided by the one-cycle 
period $\tau_{cyc}$ for cyclic heat engines.
The power $P\equiv \dot{W}$ 
is expressed as $P=-F\dot{x}$ for 
steady-state heat engines where the time-independent external force $F$ is 
acting on its conjugate variable $x$, and as $P=W/\tau_{cyc}$ for 
cyclic heat engines. 
From the decomposition 
$\dot{\sigma}\equiv J_1X_1+J_2X_2$,  
we can define the thermodynamic force $X_1\equiv F/T_c$ and its
conjugate thermodynamic 
flux $J_1\equiv \dot{x}$ for steady-state heat engines 
as well as $X_1\equiv -W/T_c$ and $J_1\equiv 1/\tau_{cyc}$ for 
cyclic heat engines.  
We can also define the other thermodynamic force 
$X_2\equiv 1/T_c-1/T_h$ and its conjugate thermodynamic 
flux $J_2\equiv \dot{Q}_h$ for both types of heat engines.
By using these thermodynamic fluxes and forces,   
the power $P$ is rewritten as $P=-J_1X_1T_c$.
We assume that these thermodynamic fluxes and forces satisfy the 
following extended Onsager relations: 
\begin{eqnarray}
&&J_1=L_{11}X_1+L_{12}X_2,\label{eq.J1}\\ 
&&J_2=L_{21}X_1+L_{22}X_2-\gamma_h {J_1}^2,\label{eq.J2} 
\end{eqnarray}   
where the features of the Onsager coefficients  
$L_{ij}$'s in eqs.~(\ref{eq.onsagerJ1}) 
and (\ref{eq.onsagerJ2}) 
are assumed to hold also in eqs.~(\ref{eq.J1}) and (\ref{eq.J2}).
In eq.~(\ref{eq.J2}),  
the new nonlinear term $-\gamma_h          
{J_1}^2$ is introduced into the standard Onsager relation 
eq.~(\ref{eq.onsagerJ2}) and $\gamma_h$ is  
assumed to be a positive constant as $\gamma_h >0$.  
We also assume that no other higher nonlinear terms arise in eqs.~(\ref{eq.J1}) and (\ref{eq.J2}) 
by considering that the coefficients of higher nonlinear terms are too small to have effective 
contributions to $\dot{\sigma}$ compared to $-\gamma_h {J_1}^2$, as will be seen in eq.~(\ref{eq.sigma}).
In our model, 
we also assume that $X_1$ and $X_2$ are not restricted to small values, 
unlike in the linear irreversible heat engine described by 
eqs.~(\ref{eq.onsagerJ1}) and (\ref{eq.onsagerJ2}), where the limit of 
$X_1 \to 0$ and $X_2 \to 0$ is taken.
\begin{figure}
\onefigure[scale=0.27]{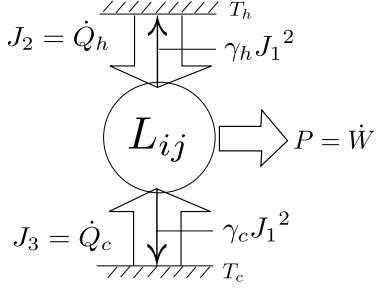}
\caption{Schematic illustration of the minimally nonlinear irreversible
heat engine described by eqs.~(\ref{eq.J2-2}) and (\ref{eq.J3-2}).}\label{fig.engine}
\end{figure}

Here we introduce the heat flux from the cold heat reservoir
\begin{eqnarray} 
\dot{Q}_c=P-\dot{Q}_h=-J_1X_1T_c-J_2\equiv J_3.\label{eq.J3}
\end{eqnarray}
Then we can rewrite the heat fluxes  
eqs.~(\ref{eq.J2}) and (\ref{eq.J3})                 
into more suggestive and symmetrical forms as (see
fig.~\ref{fig.engine}) 
\begin{eqnarray} 
&&J_2=\frac{L_{21}}{L_{11}}J_1+L_{22}(1-q^2)X_2-\gamma_h 
 {J_1}^2,\label{eq.J2-2}\\  
&&J_3=-\frac{L_{21}T_c}{L_{11}T_h}J_1-L_{22}(1-q^2)X_2-\gamma_c {J_1}^2,
 \label{eq.J3-2}  
\end{eqnarray} 
where $q$ 
is defined in eq.~(\ref{eq.csp})  
and $\gamma_c$ is defined as 
\begin{eqnarray} 
\gamma_c\equiv \frac{T_c}{L_{11}}-\gamma_h.\label{eq.gammac}  
\end{eqnarray} 
We assume $\gamma_c$ to be a positive constant as $\gamma_c>0$.
We can use eqs.~(\ref{eq.J2-2}) and (\ref{eq.J3-2}) to describe the heat
engines instead of eqs.~(\ref{eq.J1}) and (\ref{eq.J2}), regarding 
$J_1$ as the control parameter of the heat engines instead of $X_1$
since $J_1$ and $X_1$ 
are uniquely related through eq.~(\ref{eq.J1}) when $X_2$ is fixed.
We call the heat engines described by eqs.~(\ref{eq.J2-2})
and (\ref{eq.J3-2}) (or the extended Onsager relations 
eqs.~(\ref{eq.J1}) and (\ref{eq.J2})) the minimally
nonlinear irreversible heat engines. 
The term ``minimally'' implies that we 
take into account only $-\gamma_h 
{J_1}^2$ and $-\gamma_c {J_1}^2$ as the nonlinear terms. 
As we explain below, those terms will turn out to be 
the inevitable power dissipations accompanied by the finite-time 
motion of the heat engines.         

The power $P=-J_1X_1T_c=J_2+J_3$   
can be rewritten as               
\begin{eqnarray}  
P=\frac{L_{21}}{L_{11}}{\eta}_C J_1-\frac{T_c}{L_{11}}{J_1}^2,\label{eq.power}
\end{eqnarray} 
by adding eqs.~(\ref{eq.J2-2}) and (\ref{eq.J3-2}). 
Here we immediately notice that 
the second terms in eqs.~(\ref{eq.J2-2}) and (\ref{eq.J3-2}) 
do not contribute to eq.~(\ref{eq.power}) at all.   
They mean just the direct 
heat transfer from the hot heat reservoir to the cold one, which
arises in the case of the non-tight coupling condition $|q|\ne 1$.
Note that similar decomposition of the power  
like eq.~(\ref{eq.power}) is also given in~\cite{GMS}.    
The first term in eq.~(\ref{eq.power}) is proportional to $\Delta T$ 
through ${\eta}_C$, 
meaning the power generation due to the temperature difference. 
On the other hand, the second term in eq.~(\ref{eq.power}) 
remains nonzero even when $\Delta T=0$
and means the power dissipation 
which should necessarily be consumed 
once the heat engine moves at a finite rate ($J_1 \ne 0$) 
regardless of how small $J_1$ is. The power dissipation results 
in the increase of the total internal energy of the heat reservoirs.
When $\Delta T=0$, this is nothing but the effect of Joule heating,
which is seen if we can consider that $T_c/L_{11}$ and $J_1$ in 
eq.~(\ref{eq.power}) correspond to resistance and electric current, 
respectively. 

To clarify the physical meaning of each term in eqs.~(\ref{eq.J2-2})
and (\ref{eq.J3-2}) in more detail,  
we rewrite the total entropy production rate 
$\dot{\sigma}=-\dot{Q}_h/T_h-\dot{Q}_c/T_c=-J_2/T_h-J_3/T_c$ as 
\begin{eqnarray} 
\dot{\sigma}=L_{22}(1-q^2){X_2}^2+\frac{{J_1}^2}{L_{11}}
-\gamma_h {J_1}^2 {X_2},\label{eq.sigma}
\end{eqnarray}
by using eqs.~(\ref{eq.J2-2}) and (\ref{eq.J3-2}).
The first term means the entropy increase rate of  
the heat reservoirs due to the direct heat transfer.
The second term comes
from the inevitable work consumption due to the finite-time operation.
The third term in eq.~(\ref{eq.sigma}) arises 
due to the presence of $-\gamma_h {J_1}^2$ in eq.~(\ref{eq.J2}). 
In the case of the linear irreversible heat engine 
described by eqs.~(\ref{eq.onsagerJ1}) 
and (\ref{eq.onsagerJ2}), 
the third term is suppressed and the
non-negativity of $\dot{\sigma}$ restricts 
the Onsager coefficients $L_{ij}$'s to eq.~(\ref{eq.restriction}).  
Even in our model, we assume 
that the restriction eq.~(\ref{eq.restriction}) still holds
although $X_1$ and $X_2$ are not restricted to small values.
Then the non-negativity of eq.~(\ref{eq.sigma}) is always 
guaranteed since it is rewritten as $\dot{\sigma}=L_{22}(1-q^2){X_2}^2
+(\gamma_h/T_h+\gamma_c/T_c){J_1}^2$ by using eq.~(\ref{eq.gammac}): 
the first term is always non-negative due to eq.~(\ref{eq.restriction}) and 
the second one is also always non-negative due to the assumptions 
$\gamma_h>0$ and $\gamma_c>0$.
\section{Efficiency at maximum power}
We consider the efficiency of the heat engine  
$\eta=W/Q_h=P/\dot{Q}_h=P/J_2$, where $P$ and $J_2$ are given 
in eqs.~(\ref{eq.power}) and (\ref{eq.J2-2}), respectively.
When $X_2$ and $L_{ij}$'s are given,  
the maximum power is realized at $J_1=L_{12}X_2/2$ as a 
solution of $\partial P/\partial J_1=0$ and  
then we obtain the maximum power $P^*$ 
and the efficiency at the maximum power $\eta^*$ as 
\begin{eqnarray} 
&&P^*=\frac{q^2L_{22}{\Delta T}^2}{4{T_h}^2T_c},\label{eq.pmax}\\
&&\eta^*=\frac{{\eta}_C}{2}
 \frac{q^2}{2-q^2\left(1+{\eta}_C/(2(1+\gamma_c/\gamma_h))\right)},\label{eq.effi-at-pmax}
\end{eqnarray}    
respectively. 
The formula eq.~(\ref{eq.effi-at-pmax})
is the main result of this paper. 
We notice that it includes the formula eq.~(\ref{eq.broecketa}) of the 
linear irreversible heat engine as the linear term of 
$\eta_C \simeq \Delta T/T$ in the limit of $\Delta T \to 0$.
We also notice that eq.~(\ref{eq.effi-at-pmax}) has
the lower bound $\eta_-^{q}$ and the upper bound $\eta_+^{q}$ at a fixed $q$ as  
\begin{eqnarray}
\eta_-^{q}\equiv 
\frac{{\eta}_C}{2}\frac{q^2}{2-q^2}
\le
\eta^*\le \frac{{\eta}_C}{2}
\frac{q^2}{2-q^2\left(1+{\eta}_C/2\right)}\equiv \eta_+^{q},\label{eq.etaq+-}
\end{eqnarray}
by taking the asymmetrical dissipation 
limits $\gamma_c/\gamma_h \to \infty$ and 
$\gamma_c/\gamma_h \to 0$, respectively. 
Moreover $\eta_+^{q}$ takes the maximum value
\begin{eqnarray}
\eta_+^q \le \frac{\eta_C}{2-\eta_C}\equiv \eta_+,\label{eq.etam}
\end{eqnarray}
when the tight-coupling condition $|q|=1$ is satisfied
in $\eta_+^{q}$. 
Therefore this $\eta_+$ is the upper bound of $\eta^*$ 
for the minimally nonlinear irreversible heat engines.

We note that $\eta_+$ has also been found in the previous studies
on the efficiency at the maximum power:
in a finite-time heat engine model~\cite{CY}, 
where the heat fluxes are assumed to obey a specific conduction law, 
$\eta_+$ arises as a limiting case. 
In a Feynman ratchet model~\cite{VRMC}, $\eta_+$ has been obtained 
as Feynman efficiency under the no heat leak condition 
between the heat reservoirs.  
In a few finite-time Carnot cycle models~\cite{SS,EKLB1,EKLB2},  
$\eta_+$ has been found as the upper bound of $\eta^*$ in the  
asymmetrical dissipation limit.   
Finally in~\cite{GMS} (see also~\cite{MGS}),
$\eta_+$ has been proved to be the upper 
bound of $\eta^*$ in the asymmetrical dissipation limit
in a general and model-independent way. But the proof  
in~\cite{GMS} is limited to the case of stochastic steady-state 
heat engines. Our theory can be applicable  
to cyclic heat engines as well and unify these previous results. 
 
Here we stress physical importance of $\eta_+$: 
if the efficiency at the maximum power of a 
finite-time heat engine exceeds $\eta_+$, it 
implies that the heat engine works under higher degree of nonequilibrium.
In fact, we can see that $\eta^*$ of
the finite-time Carnot cycle model of ideal gas reported in~\cite{IO1,IO2}
exceeds $\eta_+$ due to the higher nonequilibrium effect~\cite{YI}. 
Therefore $\eta_+$ could be a criterion for determining
the degree of nonequilibrium of finite-time heat engines.
 
\section{Example: low-dissipation Carnot engine}
For a demonstration of the validity of 
our theory, we show that the low-dissipation Carnot
engine~\cite{EKLB2} is described 
by the extended Onsager relations eqs.~(\ref{eq.J1}) and (\ref{eq.J2}).
Here the low-dissipation Carnot engine 
is a heat engine model proposed as 
a finite-time extension of the quasistatic Carnot cycle. It
assumes the specific form of the heats transferred from the heat 
reservoirs during the isothermal processes as 
\begin{eqnarray}
&&Q_h=T_h \Delta S-\frac{T_h\Sigma_h}{\tau_h}+\cdots ,\label{eq.low1}\\
&&Q_c=-T_c\Delta S-\frac{T_c\Sigma_c}{\tau_c}+\cdots ,\label{eq.low2}
\end{eqnarray}
where $\Delta S$ is the quasistatic entropy change 
inside the heat engine during the isothermal process in contact with the
hot heat reservoir, $\tau_h$ and $\tau_c$ are the durations during
the isothermal processes in contact with the hot heat reservoir 
and the cold one, respectively, and $\Sigma_h$ and $\Sigma_c$ 
are positive constants. 
We consider that the constants $\Sigma_h$ and $\Sigma_c$ contain the 
details how the engine deviates from the quasistatic limit. 
The assumption eqs.~(\ref{eq.low1}) and (\ref{eq.low2}) means that 
the lowest deviation from the quasistatic heat should be proportional 
to the inverse of the duration.   
In a stochastic finite-time Carnot cycle model~\cite{SS} analyzed by  
the Fokker-Planck equation, such a ${\tau}^{-1}$ term
indeed arises [cf. eq.~(16) in~\cite{SS}]. 
In a finite-time Carnot cycle model of ideal gas
analyzed by the molecular kinetic theory~\cite{IO1},  
it is also confirmed that the lowest deviation 
is proportional to ${\tau}^{-1}$ 
[cf. eq.~(11) in~\cite{IO1}].  
Finally such a ${\tau}^{-1}$ term also 
arises in a quantum dot Carnot engine model~\cite{EKLB1}
analyzed by the master equation approach [cf. eq.~(28) 
in~\cite{EKLB1}].
Therefore the assumption of the specific form of the heats eqs.~(\ref{eq.low1}) 
and (\ref{eq.low2}) has microscopically 
been justified in these models.
Additionally, we neglect higher order terms such as $O({\tau}^{-2})$
in eqs.~(\ref{eq.low1}) and (\ref{eq.low2}) 
in this low-dissipation approximation.

We can express the power $P=(Q_h+Q_c)/(\tau_h+\tau_c)$ of  
this engine as $P=(\Delta T \Delta                              
S-T_h\Sigma_h/\tau_h-T_c\Sigma_c/\tau_c)/(\tau_h+\tau_c)$ by using  
eqs.~(\ref{eq.low1}) and (\ref{eq.low2}). Maximizing this power  
by the durations $\tau_h$ and $\tau_c$ as $\partial P/\partial 
\tau_h=\partial P/\partial \tau_c=0$, we find the physically relevant  
solutions as
\begin{eqnarray}
&&\tau_h=\frac{2T_h \Sigma_h}{(T_h-T_c)\Delta S}\left(1+\sqrt{\frac{T_c
						 \Sigma_c}{T_h 
						 \Sigma_h}}\right)\equiv
{\tau_h}^*,\label{eq.low.tauh}\\
&&\tau_c=\frac{2T_c \Sigma_c}{(T_h-T_c)\Delta S}\left(1+\sqrt{\frac{T_h
						 \Sigma_h}{T_c 
						 \Sigma_c}}\right)\equiv
{\tau_c}^*.\label{eq.low.tauc}  
\end{eqnarray}   
Then by using the definition $\eta=(Q_h+Q_c)/Q_h$ 
and eqs.~(\ref{eq.low1}), (\ref{eq.low2}), 
(\ref{eq.low.tauh}) and (\ref{eq.low.tauc}), 
we can obtain the efficiency at the maximum power $\eta^*$ as
\begin{eqnarray}
\eta^*=\frac{\eta_C\left(1+\sqrt{\frac{T_c\Sigma_c}{T_h\Sigma_h}}\right)}{\left(1+\sqrt{\frac{T_c\Sigma_c}{T_h\Sigma_h}}\right)^2+\left(1-\frac{\Sigma_c}{\Sigma_h}\right)\frac{T_c}{T_h}}.
\label{eq.low-etapmax} 
\end{eqnarray} 
We can easily notice that 
eq.~(\ref{eq.low-etapmax}) is bounded from the lower side and the 
upper side as
\begin{eqnarray}   
\frac{\eta_C}{2}\le \eta^* \le \frac{\eta_C}{2-\eta_C},\label{eq.low+-}  
\end{eqnarray}   
by taking the asymmetrical dissipation limits $\Sigma_c/\Sigma_h \to \infty$   
and $\Sigma_c/\Sigma_h \to 0$, respectively~\cite{EKLB2}    
(see also~\cite{EKLB1} for    
the derivation of these bounds in a quantum dot Carnot engine model).   
In~\cite{EKLB2}, it is stated that observed efficiencies of various
actual power plants tend to locate between these two bounds.
It is also interesting to see that the same bounds 
were derived in a different finite-time heat engine model
based on a specific heat conduction law~\cite{CY}. 
By comparing eq.~(\ref{eq.low+-}) with eq.~(\ref{eq.etaq+-}), 
we may consider that the tight-coupling condition 
$|q|=1$ holds in this low-dissipation Carnot engine. 
We can prove it by writing the extended Onsager relations of this engine 
explicitly as follows.
  
First, we consider the total entropy production rate 
\begin{eqnarray}
\dot{\sigma}&=&-\frac{\dot{Q}_{h}}{T_{h}}-\frac{\dot{Q}_{c}}{T_{c}}= 
-\frac{\dot{W}}{T_c}+\dot{Q}_h \left(\frac{1}{T_c}-\frac{1}{T_h}\right)\nonumber\\ 
&=&-\frac{W}{T_c(\alpha+1)\tau_h}+\dot{Q}_h \left(\frac{1}{T_c}-\frac{1}{T_h}\right),\label{eq.sigma_low}
\end{eqnarray}
where we have defined the parameter $\alpha$ as $\alpha\equiv
\tau_c/\tau_h$ and the dot denotes the quantity divided by the one-cycle
period $\tau_{cyc}=\tau_h+\tau_c=(\alpha+1)\tau_h$.  
From the decomposition  
$\dot{\sigma}=J_1X_1+J_2X_2$, 
we can define the thermodynamic forces  
$X_1\equiv -W/T_c$, $X_2\equiv 1/T_c-1/T_h$ and their  
corresponding thermodynamic fluxes $J_1\equiv 1/((\alpha+1)\tau_h)$,   
$J_2\equiv \dot{Q}_h$. 
Using eqs.~(\ref{eq.low1}), (\ref{eq.low2}) 
and the definitions of the thermodynamic forces and fluxes,
we can easily calculate the Onsager coefficients $L_{ij}$'s and the constant
$\gamma_h$ of this low-dissipation
Carnot engine as 
\begin{eqnarray}  
&&L_{11}=\frac{T_c}{(T_h\Sigma_{h}+T_c\Sigma_{c}/\alpha)(\alpha+1)},\label{eq.minimal-L11}\\
&&L_{12}=\frac{T_hT_c\Delta S}{(T_h\Sigma_{h}+T_c\Sigma_{c}/\alpha)(\alpha+1)},\label{eq.minimal-L12}\\
&&L_{21}=\frac{T_hT_c\Delta S}{(T_h\Sigma_{h}+T_c\Sigma_{c}/\alpha)(\alpha+1)},\label{eq.minimal-L21}\\
&&L_{22}=\frac{{T_h}^2T_c{\Delta
 S}^2}{(T_h\Sigma_{h}+T_c\Sigma_{c}/\alpha)(\alpha+1)},\label{eq.minimal-L22}\\  
&&\gamma_h=T_h\Sigma_h(\alpha+1),\label{eq.minimal-gammah}
\end{eqnarray}
respectively. 
$\gamma_c$ is also given by 
\begin{eqnarray}
\gamma_c=\frac{T_c\Sigma_{c}(\alpha+1)}{\alpha},\label{eq.minimal-gammac}  
\end{eqnarray}                                                        
by using eqs.~(\ref{eq.minimal-L11}), (\ref{eq.minimal-gammah}) and (\ref{eq.gammac}).
We notice that the reciprocity $L_{12}=L_{21}$ surely holds from eqs.~(\ref{eq.minimal-L12}) and (\ref{eq.minimal-L21}). 
Moreover we can confirm the tight-coupling condition 
$|q|=|L_{12}/\sqrt{L_{11}L_{22}}|=1$ from eqs.~(\ref{eq.minimal-L11}),
(\ref{eq.minimal-L12}) and (\ref{eq.minimal-L22}) as expected. 
We can obtain $\eta^*$ of the low-dissipation Carnot engine as 
\begin{eqnarray}
&&\eta^*=\frac{{\eta}_C}{2}
 \frac{1}{2-\left(1+{\eta}_C/\left(2\left(1+\frac{T_c\Sigma_c}{\alpha T_h\Sigma_h}\right)\right)\right)},\label{eq.effi-at-pmax2}
\end{eqnarray}
by substituting $|q|=1$, eqs.~(\ref{eq.minimal-gammah}) and 
(\ref{eq.minimal-gammac}) into eq.~(\ref{eq.effi-at-pmax}).
We can also obtain $P^*$ as 
\begin{eqnarray}
P^*=\frac{{\Delta
 S}^2 {\Delta T}^2}
{4(T_h\Sigma_{h}+T_c\Sigma_{c}/\alpha)(\alpha+1)},\label{eq.pmax2}
\end{eqnarray}     
by substituting $|q|=1$, eq.~(\ref{eq.minimal-L22}) into eq.~(\ref{eq.pmax}).
However we notice that eq.~(\ref{eq.pmax2}) 
still contains the tunable parameter $\alpha$ 
and can further be maximized as $\partial 
P^*(\alpha)/\partial \alpha=0$, which reduces to  
$\alpha=\sqrt{T_c\Sigma_c/(T_h\Sigma_h)}\equiv {\alpha}^*$.
From eqs.~(\ref{eq.low.tauh}) and (\ref{eq.low.tauc}),
we can see that ${\alpha}^*={\tau_c}^*/{\tau_h}^*$ holds.   
Substituting this ${\alpha}^*$ 
into eq.~(\ref{eq.effi-at-pmax2}), 
we finally reproduce
eq.~(\ref{eq.low-etapmax}).                                 
Therefore we can conclude that the low-dissipation                        
Carnot engine is exactly described by the extended Onsager relations.
In other words, the inclusion of the power dissipation term $-\gamma_h {J_1}^2$
into the Onsager relation as in eq.~(\ref{eq.J2}) is justified by this
explicit example, whose assumptions eqs.~(\ref{eq.low1}) and (\ref{eq.low2}) 
are consistent with the microscopically analyzed models~\cite{SS,EKLB1}.
  
\section{Summary and discussion}
We proposed the minimally nonlinear irreversible heat 
engine described by the extended Onsager relations,  
where a new nonlinear term meaning the power dissipation  
is added to the heat flux from the hot heat reservoir 
in the standard Onsager relation     
and no other nonlinear terms are assumed to arise.  
Thus our model can be regarded as a natural and minimal extension 
of the linear irreversible heat engine.
We formulated the efficiency at the maximum power $\eta^*$  
of our model and showed that it is bounded from the upper side 
by $\eta_C/(2-\eta_C)$. This upper bound can be attained when the 
heat engine satisfies the tight-coupling condition $|q|=1$ and the asymmetrical
dissipation limit $\gamma_c/\gamma_h \to 0$ is taken.  
As a demonstration of the validity of our theory, 
we explicitly wrote down the extended Onsager relations 
of the low-dissipation Carnot engine~\cite{EKLB2} and confirmed that 
it satisfies the tight-coupling condition $|q|=1$. 
Though the low-dissipation Carnot engine is an example of the 
cyclic heat engine, we should note that the power dissipation terms arise
also in a few steady-state systems~\cite{VBKM,VBK,VBVB}, where  
analytical calculations of the Onsager coefficients $L_{ij}$'s, $\gamma_h$ and 
$\gamma_c$ are explicitly done based on a molecular kinetic theory. 
These calculations and the present example of the  
low-dissipation Carnot engine in this paper  
could support the validity of our theory,   
which treats the cyclic heat engines and the steady-state ones 
in the unified manner.
It will be a future challenge to find    
the upper bound of the efficiency at the maximum power for more general  
irreversible heat engines with higher nonlinear terms beyond our model. 

\acknowledgments
The authors thank N. Ito, K. Nemoto, M. Hoshina and S. Oono for helpful 
discussions. 
The authors also thank the Yukawa Institute for Theoretical
Physics at Kyoto University.  
Discussions during the YITP workshop YITP-W-10-16 on 
"YITP Workshop 2010: Physics of Nonequilibrium Systems -Fluctuation 
and Collective Behavior-" 
were useful to complete this work.  
YI acknowledges the financial support from a
Grant-in-Aid for JSPS Fellows (Grant No. 22-2109).

\end{document}